\documentclass[]{spie}  

\usepackage{amsmath,amsfonts,amssymb}
\usepackage{graphicx}
\usepackage[colorlinks=true, allcolors=blue]{hyperref}
\usepackage{float}

\title{3D printing for astronomical mirrors}

\author[a]{Melanie Roulet}
\author[b]{Carolyn Atkins}
\author[a]{Emmanuel Hugot}
\author[a]{Sabri Lemared}
\author[a]{Simona Lombardo}
\author[a]{Marc Ferrari}

\affil[a]{Aix Marseille Univ, CNRS, CNES, LAM, Marseille, France}
\affil[b]{UK Astronomy Technology Centre, Royal Observatory, Edinburgh, EH9 3HJ, UK}

\authorinfo{melanie.roulet@lam.fr}

\begin{document} 
\maketitle

\begin{abstract}
3D printing, also called additive manufacturing, offers a new vision for optical fabrication in term of achievable optical quality and reduction of weight and cost. In this paper we describe two different ways to use this technique in the fabrication process. The first method makes use of 3D printing in the fabrication of warping harnesses for stress polishing, and we apply that to the fabrication of the WFIRST coronagraph off axis parabolas. The second method considers a proof of concept for 3D printing of lightweight X-Ray mirrors, targeting the next generation of X-rays telescopes. Stress polishing is well suited for the fabrication of the high quality off axis parabolas required by the coronagraph to image exoplanets.. Here we describe a new design of warping harness which can generate astigmatism and coma with only one actuator. The idea is to incorporate 3D printing in the manufacturing of the warping harness. The method depicted in this paper demonstrates that we reach the tight precision required at the mirrors’ surface. Moreover the error introduced by the warping harness fabricated by 3D printing does not impact the final error budget. Concerning the proof of concept project, we investigate 3D printing towards lightweight X-ray mirrors. We present the surface metrology of test samples fabricated by stereo lithography (SLA) and Selective Laser Sintering (SLS) with different materials. The lightweighting of the samples is composed of a series of arches. By complementing 3D printing with finite element analysis topology optimization we can simulate a specific optimum shape for the given input parameters and external boundary conditions. The next set of prototypes is designed taking to account the calculation of topology optimisation.   
\end{abstract}


\section{INTRODUCTION}
\label{sec:intro}  

3D printing could revolutionise the manufacture of optics and telescope structures. This new manufacturing process was initiated in the early 1980s, \cite{Jane2012} first used with plastics the technology offers now a wide range of materials from ceramics to metals as well as composite materials. Many studies started to investigate 3D printed optics both for space and ground telescopes and the results are very promising \cite{Enrico2018} \cite{SPIEpaper}. In this paper we focus on space optics. The main requirements regarding the launch phase for space optics are their rigidity and their mass. 3D printing could provide lightweight structures which are not fabricable with traditional mechanical manufacturing; therefore the scope for mirror and lightweighting design has become substantially wider.Combined with topology optimisation, this technology is even more powerful to create the optimum shape for the given requirements. 3D printing could also be combined with polishing processes, like stress polishing for the creation of free-forms super-smooth optics. In Section 1 is described the design phase of a printed warping harness for the manufacturing of the WFIRST off axis parabolas. In this case 3D printing is used to create innovative design for the warping harness able to generate a force distribution corresponding to astigmatism and coma deformations on the substrate. In Section 2, direct 3D printed substrates were investigated. This study project aimed to explore the surface quality of the printed samples after printing and with simple post processing, compare the different materials and their behaviour after printing.

\begin{figure}[H]
  \centering
  \includegraphics[width = 0.5\linewidth]{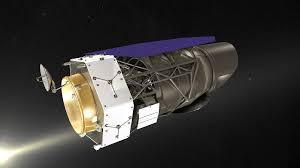}\hfill
  \includegraphics[width = 0.5\linewidth]{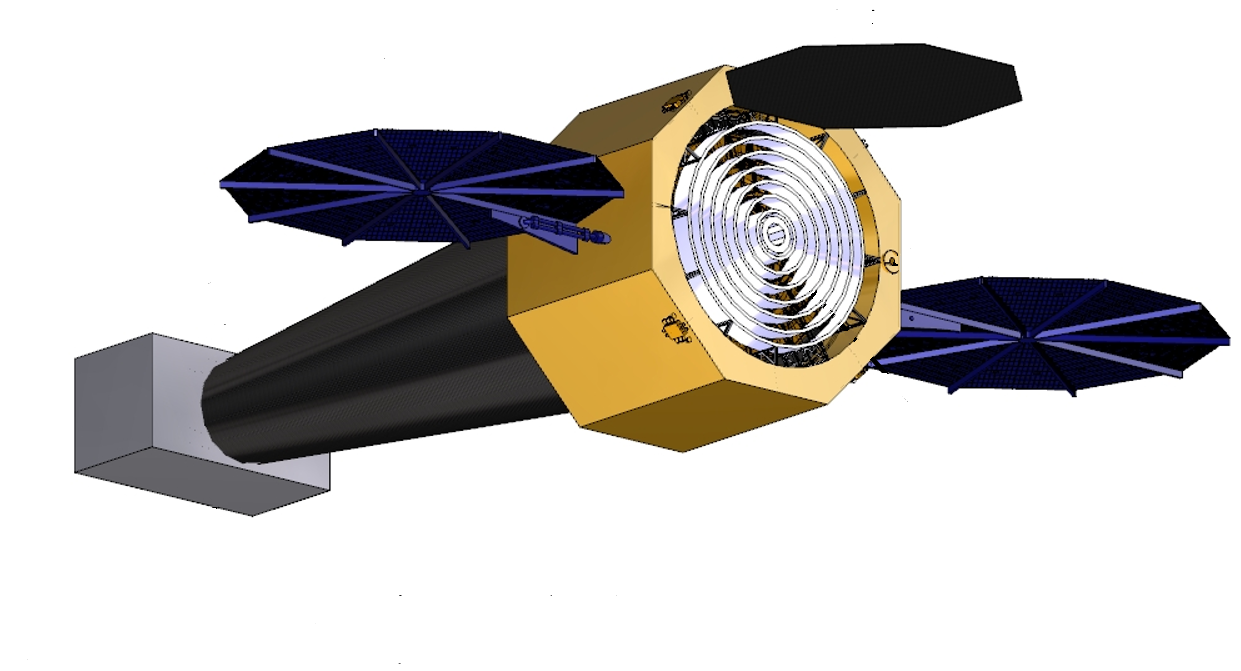}
  \caption{WFIRST telescope (left) Lynx telescope (right) (credit: NASA)}
  \label{intro}
\end{figure}

In the first part of this paper we focus on the Wide Field Infrared Survey Telescope (WFIRST)\cite{2015arXiv150303757S} working in the infrared, in Figure \ref{intro}(left). We focus on the development of a stress polishing method for the mirrors of the coronagraphic instrument. These mirrors, off axis parabolas, must have an high surface quality in order to provide the minimal amount of errors and then not degrade the final image sharpness. More specifically, mid and high spatial frequency ripples must be avoided on the surfaces, at the scale of the frequency were the instrument is seeking planets. 
For that purpose, we use the stress polishing method, which principle consists in applying a spherical polishing on a warped substrate, in order to imprint the shape of the warping applied onto the reflective surface. The idea is to create printed harness to imprint astigmatism and coma deformation. The use of 3D printing allows using complex shape structure and also lightweight structure for the harness to generate non symmetrical force distribution. 

In a second part of this study, we focus on X-ray field and the Lynx design concept, in Figure \ref{intro}(right)\cite{reflynx}. X-ray needs extreme precision optics and optimum mass \cite{Stephen2016}. We use 3D printing to investigate an alternative method to create high surface quality lightweight mirrors. At the same time using topology optimisation we can reduce the weight and the bulkiness. In this way we compared the 3D printing process and the material and we studied the feasibility of lightweight structures manufactured by 3D printing. We investigated surface quality and roughness after the printing and after applying post-processing. 

\section{Off axis parabola}
WFIRST coronagraph is the case study to develop a new fabrication process for off axis parabola. This instrument uses eight off axis parabola as relay optics. Additive manufacturing allows to think about innovative designs for the warping harness, with hollow structure or innovative shape. \cite{2015arXiv150303757S}

\subsection{Zernike decomposition}
The decomposition of the surface shape have been done with Zernike polynomials \cite{Noll:76}. The aberrations are sorted with  an increasing radial order $n$ and a decreasing azimuthal order $m$, as shown in Table \ref{coeff}.

\begin{table}[H]
\begin{center}
\begin{tabular}{|c|c|c|c|c|}
  \hline
  Number & Aberration & Expression & n & m\\
  \hline
  1 & Piston & 1 & 0 & 0\\
  2 & Tilt x & $\rho \cos\theta$ & 1 & 1\\
  3 & Tilt x & $\rho \sin\theta$ & 1 & -1\\
  4 & Spherical & $2\rho^2 -1$ & 2 & 0\\
  5 & Astigmatism 3$x$ & $ \rho^2 \cos (2\theta)$ & 2 & 2\\
  6 & Astigmatism 3$y$ & $ \rho^2 \sin (2\theta)$ & 2 & -2\\
  7 & Coma 3$x$ & $(3\rho^2 -2)\rho \cos\theta$ & 3 & 1\\
  8 & Coma 3$y$ & $(3\rho^2 -2)\rho \sin\theta$ & 3 & -1\\
  9 & Trefoil 5$x$ & $ \rho^3 \cos (3\theta)$ & 3 & 3\\
  10 & Trefoil 5$y$ & $ \rho^3 \sin (3\theta)$ & 3 & -3\\
  11 & Spherical 3 & $6\rho^4 - 6\rho^2 + 1$ & 4 & 0\\
  12 & Astigmatism 5 $x$ & $(4\rho^2 -3)\rho^2 \cos(2\theta)$ & 4 & 2\\
  13 & Astigmatism 5 $y$ & $(4\rho^2 -3)\rho^2 \sin(2\theta)$ & 4 & -2\\
  14 & Squad 7 $x$ & $\rho^4 \cos (4\theta)$ & 4 & 4\\
  15 & Squad 7 $y$ & $\rho^4 \sin (4\theta)$ & 4 & -4\\
  16 & Coma 5 $x$ & $(10\rho^4 -12\rho^2 +3)\rho \cos\theta$ & 5 & 1\\
  17 & Coma 5 $y$ & $(10\rho^4 -12\rho^2 +3)\rho \sin\theta$& 5 & -1\\

  \hline
\end{tabular}
\caption{Zernike Aberrations}
  \label{coeff}
  \end{center}
\end{table}

\subsection{Stress polishing}

\begin{figure}[H]
  \centering
  \includegraphics[width= 0.3\linewidth]{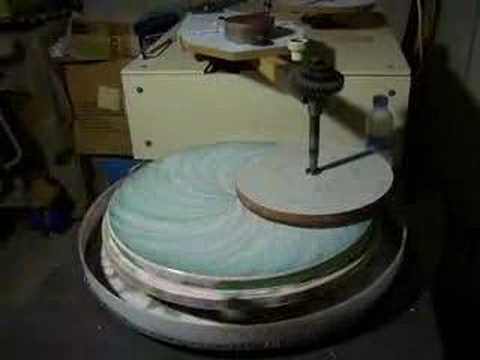}\hfill
  \includegraphics[width= 0.4\linewidth]{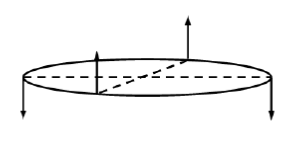}\hfill
    \includegraphics[width= 0.3\linewidth]{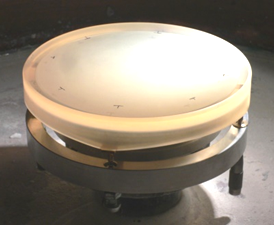}
  \caption{Picture of grinding during polishing (left), Schematic of the applied force in case of astigmatism deformation (middle), Mirror on its deformation system (right) (credit: Hugot 2008 \cite{Hugot2008})}
  \label{stressforces}
\end{figure}

Stress polishing is a polishing technique developed by the German astronomer Bernhard Schmidt in the 1930's \cite{Lemaitre:72}. The process consists in applying forces through a warping harness on the substrate during the polishing (cf Figure \ref{stressforces} (left)). Then, after the polishing process, (cf Figure \ref{stressforces} (right)) the warping harness is removed and the substrate comes back to its initial position. The warping harness must generate a warping function equal to the inverse of the required final shape. With this technique we can manufacture low cost - high performance optics, as long as the mechanical deformation is handled and precisely controlled, and within the breakage limit of the mirrors substrates. Finite element optimisation is crucial to reach the required quality. \cite{Lemaitre:72} \cite{lemaitre2008astronomical}

\subsection{Astigmatism mirror}

To manufacture astigmatism shape we use the combination of forces drawn in Figure \ref{stressforces} (middle). Two pairs of equal and opposite forces are applied on a ring underneath the substrate. This combination gives a variation of the radius of curvature in two orthogonal directions which is defined by the Equation \ref{eq}. The first mode of this equation gives a terms in $ \rho^2 \cos (2\theta)$ which corresponds to astigmatism 3 in Table \ref{coeff}. \cite{lemaitre2008astronomical}

\begin{equation}
\omega(\rho,\theta) \propto \sum_{m=2,6,10,...}^{\infty} \left ( \frac{1}{m(m-1)} + \frac{2(1+v)}{(1-v)(m-1)m^2} - \frac{\rho^2}{m(m+1)}  \right ) \rho^m \cos (m\theta)
\label{eq}
\end{equation}

\subsubsection{Simulation and Result}

\begin{figure}[H]
  \centering
  \includegraphics[height =4.5cm]{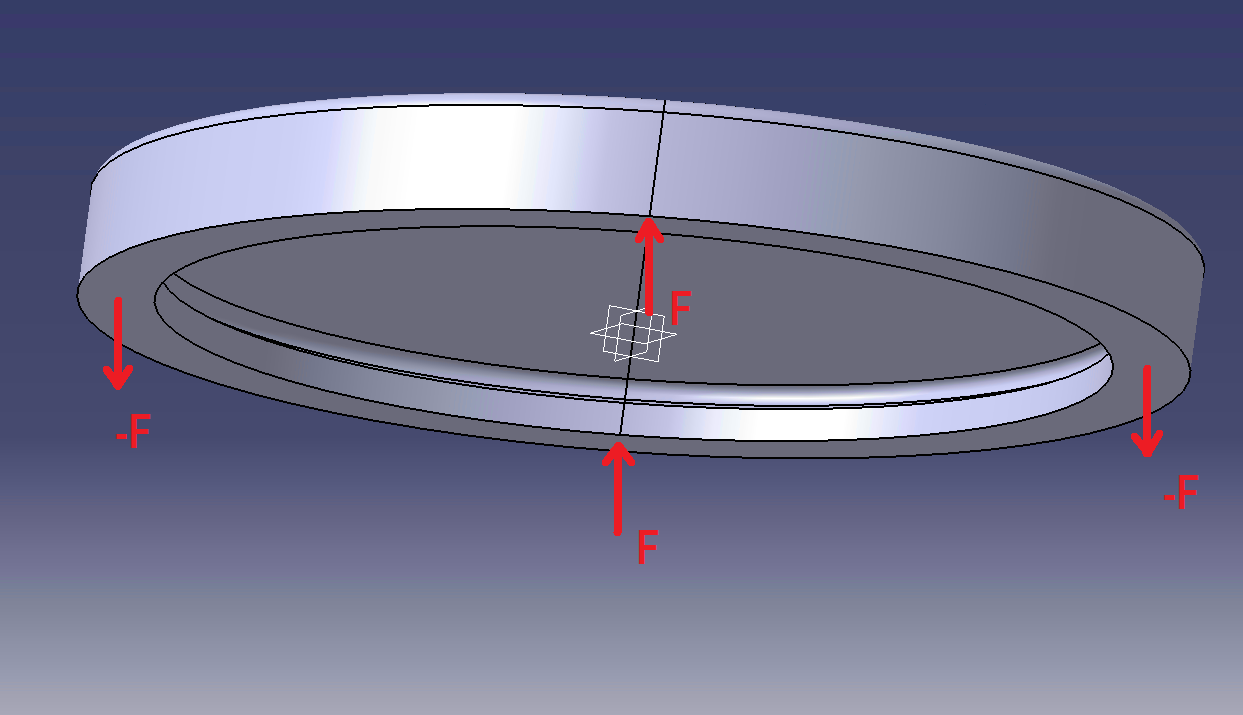}\hfill
    \includegraphics[height =4.5cm]{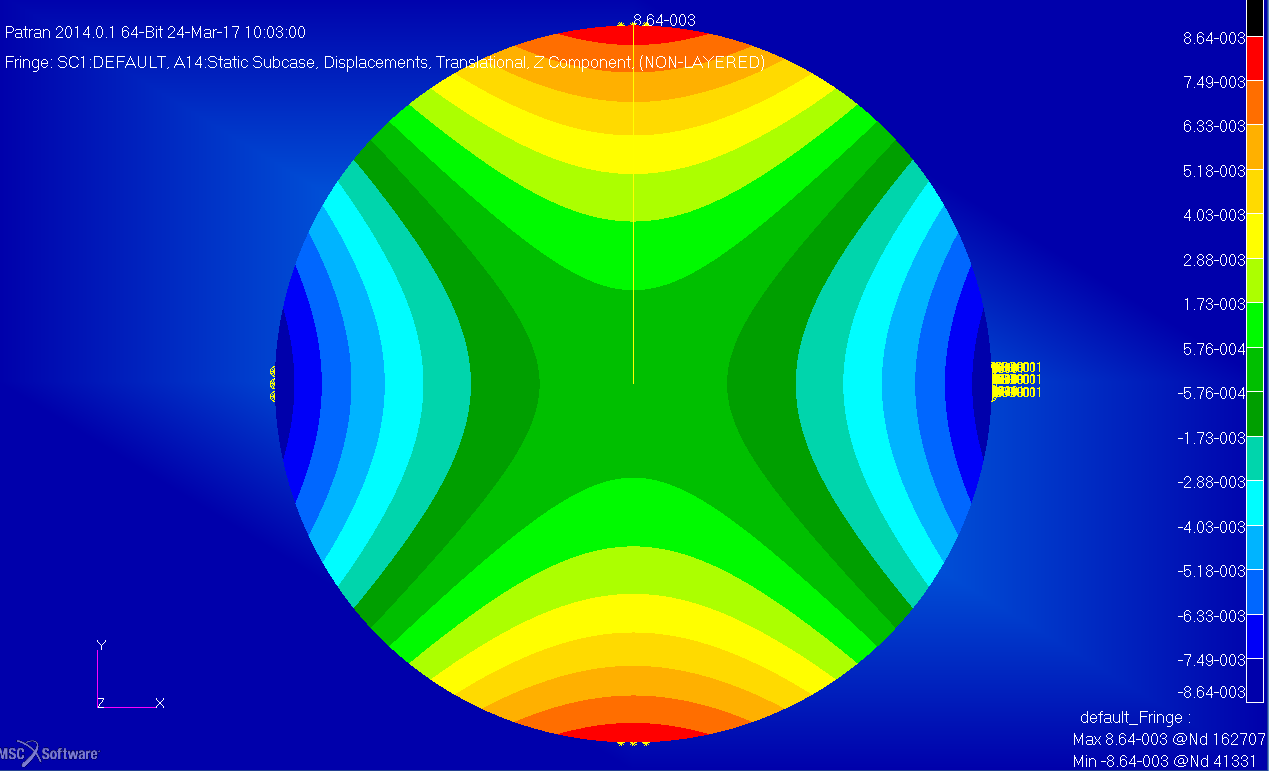}
  \caption{CAD view of the mirror with the force positions (left), Top view of the displacement after the simulation [mm] (right)}
  \label{astig}
\end{figure}

The substrate was simulated with zero expansion glass ceramic material, commonly known as Zerodur$\circledR$. This material has great capacities of polishing so we can obtain extremely smooth surface with residual roughness lower than 1 nm RMS. Additionally Zerodur$\circledR$ is often used in space application due to its very low CTE, 0 +/- 0.02 10-6/K. \cite{Schott263}
For modelling the combination of forces in Figure \ref{stressforces} (middle) we applied four forces on the ring: Two opposite forces in the negative vertical direction and perpendicularly two forces on the positive vertical direction in Figure \ref{astig} (left). 

From Patran/Nastran analysis, the node displacements in z-axis direction are shown in Figure \ref{astig} (right). The node displacement are symmetrical both on x-axis and y-axis. Comparing this simulation to the one which has been done in the PhD thesis by E.Hugot \cite{hugot:tel-00519452} we obtained as expected displacement in microns. The aberrations decomposition displayed pure astigmatism 3, Zernike numbers 5 and 6 in Table \ref{coeff} with no residuals aberrations.

\subsection{Off axis parabola}

An off axis parabola (OAP) is an optical surface which can be defined by a combination of two Zernikes aberrations added to a parent spherical surface: astigmatism and coma. As this is the combination of a even aberration (astig 3) and an odd aberration (coma 3), the idea to generate this surface shape is to break the symmetry of the warping harness. Nevertheless breaking symmetry is not enough to manufacture pure astigmatism 3 and coma 3, the design also generates unwanted aberrations, like trefoil 5. 
In this design, five optimisation parameters are considered. We carried out a set of simulations in order to delete the trefoil 5 aberrations. 

For the simulation we use the same conditions as previously, which includes Zerodur material, forces underneath the mirror's edges and centre fixed for boundary condition. We do the simulation with approximately 100,000 nodes and the meshing uses hexahedral element. We perform the simulation using the dimension of the blank test NASA model, with a diameter of 70mm. 

\begin{figure}[H]
  \centering
  \includegraphics[height =4.3cm]{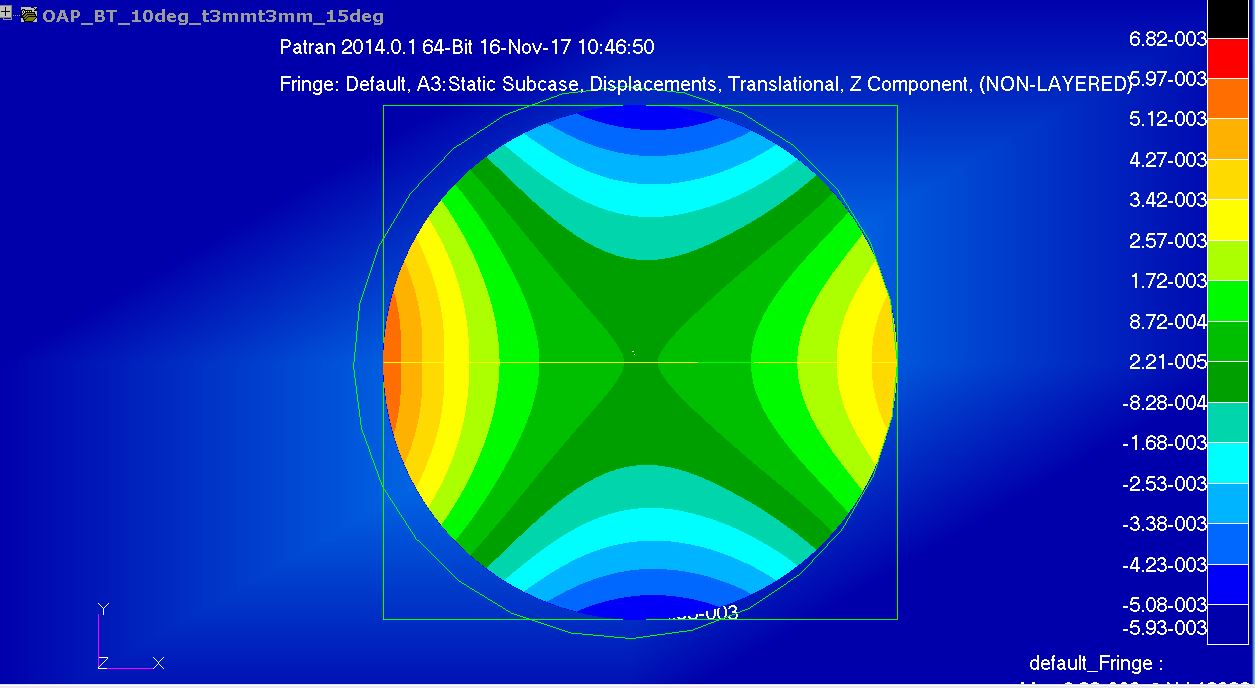}\hfill
  \includegraphics[height =4.7cm]{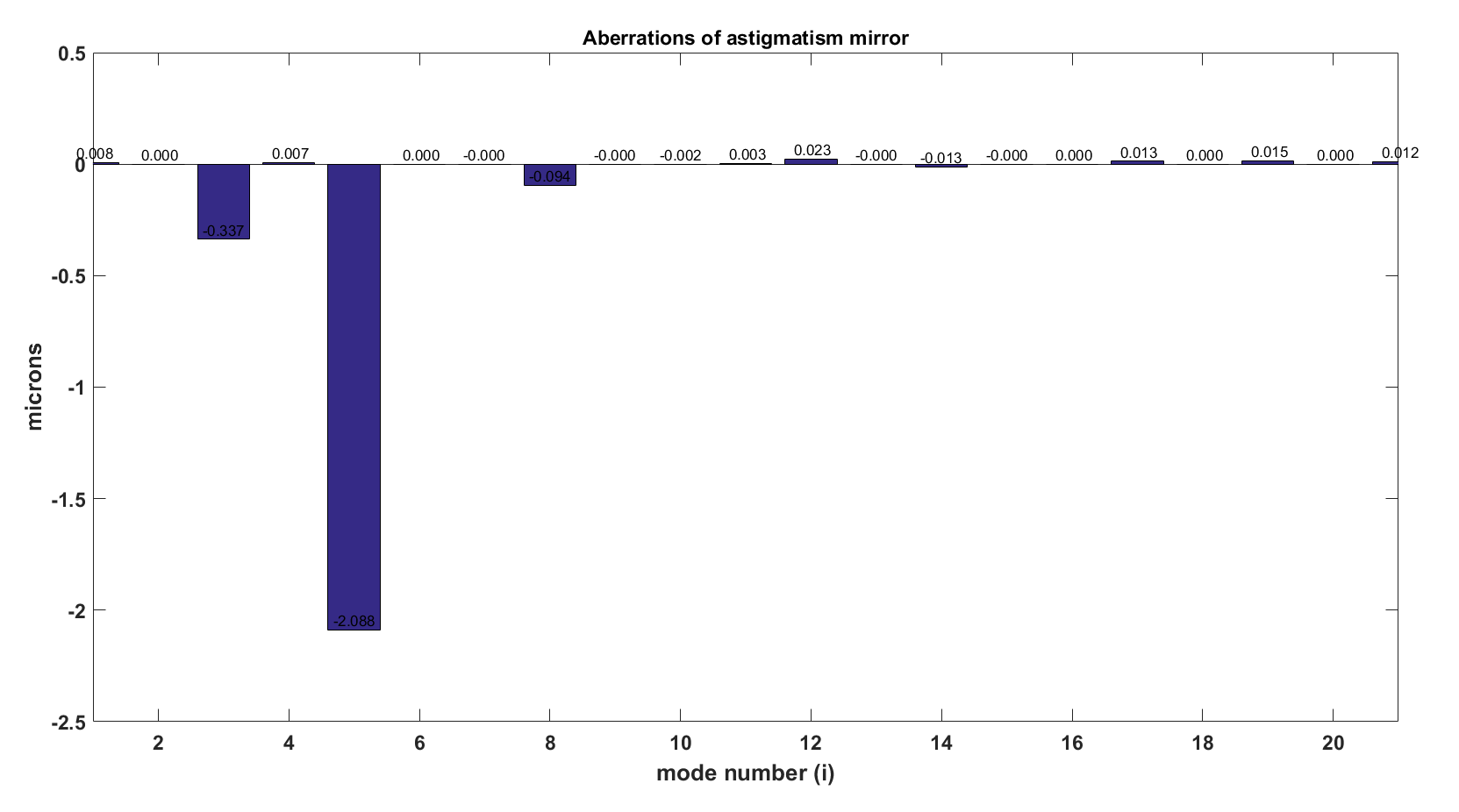}
  \caption{Top view of the displacement after the simulation [mm] (left), Zernikes decomposition in Matlab (right)}
  \label{oap}
\end{figure}

We converged on functional point where the trefoil 5 is cancelled out and the required design parameters are respected.

Now the challenge is to control the ratio between astigmatism and coma to cover all the optical requirement for the WFIRST off axis parabola. 

\subsection{Discussion}

With parametric study we find a functional point for the blank test mirror, where the warping harness generates pure astigmatism and coma. The challenging part in this study was to respect the required magnitude of astigmatism and coma and remove the residuals and also to respect the dimensions (diameter, thickness). After testing several design we were able to converge towards the solution which respects all the previous requirements. 

The next step will be to carry out the same simulation and parametric study for the eight off axis parabolas respecting their own magnitudes and dimensions. 

In the same time, we should verify our simulation data by the creation of the first prototype. Then the measurement of the fringes pattern with the interferometer will confirm or not our simulations and so our warping harness design. Then we could go further to the design of the next off axis parabola models.


\section{Lightweight structure}

In this section, we were interested in the weight and cost constraints which apply on mirrors for space telescope mission as Lynx design concept, keeping the high precision of the structure. We are interested into 3D printing and lightweight structures to create new lightweight design for future space optics.

Mirrors for X-ray telescopes were our study case, the goal was to investigate the ability to fabricate a high quality mirror surface like the precision Chandra X-ray Observatory mirrors \cite{Chandra2012}, but to use 3D printing to develop a lightweighted mirror to reduce weight towards the next generation of high spatial resolution X-ray telescope, Lynx.

\subsection{Prototype simulation}
\begin{figure}[H]
  \centering
  \includegraphics[height = 3.6cm]{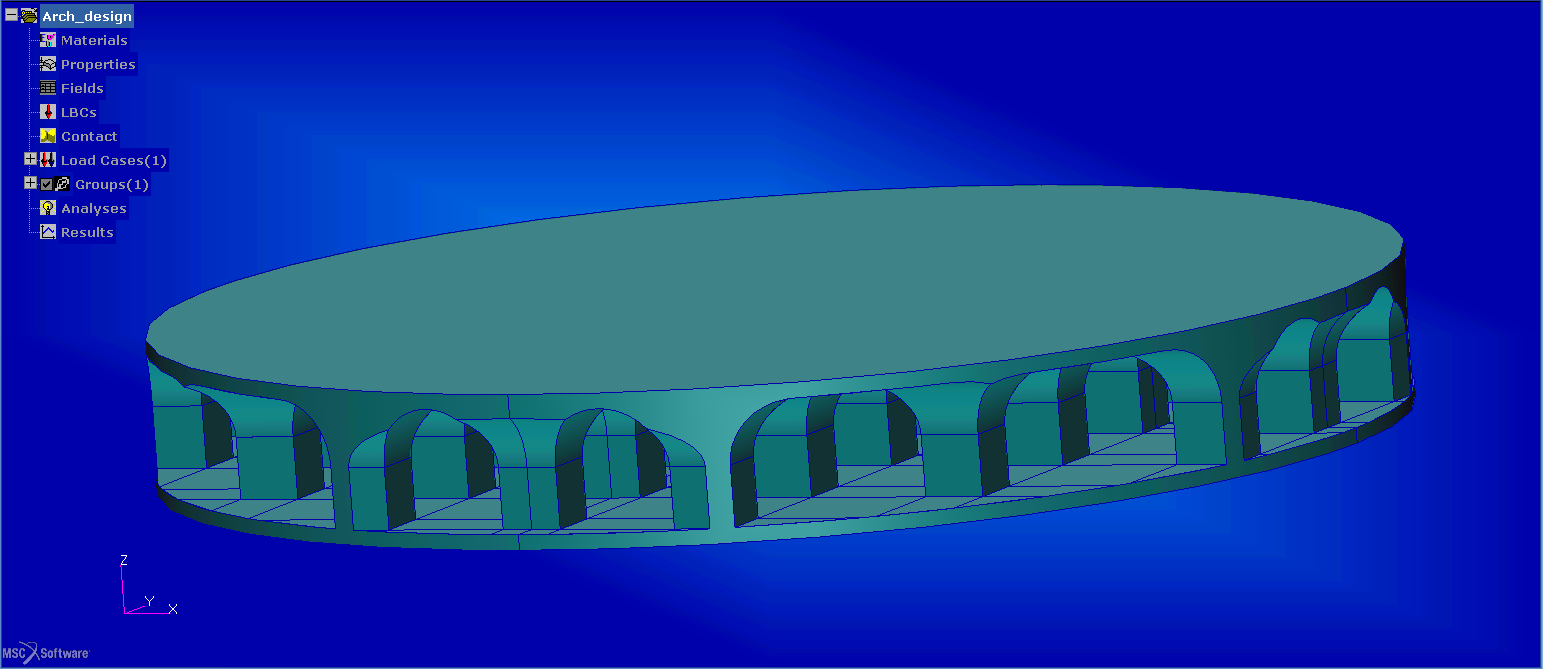}\hfill
  \includegraphics[height = 3.6cm]{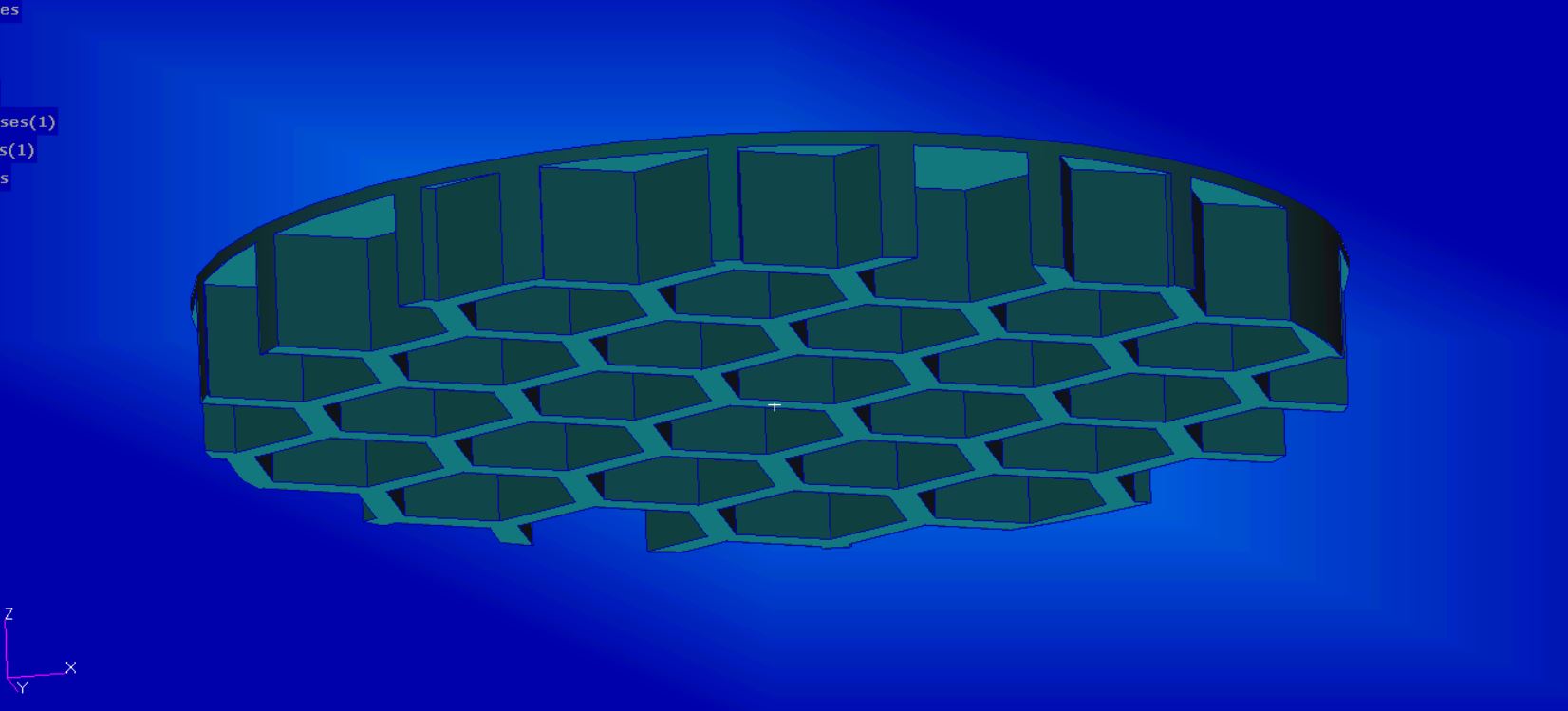}
  \caption{Arch design CAD view (left) Honeycomb design CAD view (right)}
  \label{design}
\end{figure}

In this first section is developed the comparison of two lightweight designs. One traditional lightweight structure, the honeycomb design and one new, the arch design. Honeycomb design is a common pattern for lightweight structures, it can be manufactured with mechanical tools contrary to the arch design, which can only be manufactured by 3D printing, in Figure \ref{design}.

We will compare the properties of a mechanical manufacturing lightweight design (honeycomb) with a 3D printed lightweight design (arches). Both designs were tested with finite element analysis. The prototypes are circular with a diameter of 40mm and 6mm thickness. We chose this configuration, far from the shape of X-ray mirror because polishing of circular flat surfaces is simpler than curved and rectangular surface.
Both design had 65\% mass reduction compare to a full cylinder of 6mm thickness.  The honeycomb model had no base while the arch design had a base, that doesn't impact the result on the nodes displacement. A pressure of 3500Pa, representing the polishing pressure, was applied on the reflecting surface, the base was fixed and the material is aluminium.

\begin{figure}[H]
  \centering
  \includegraphics[height = 4.7cm]{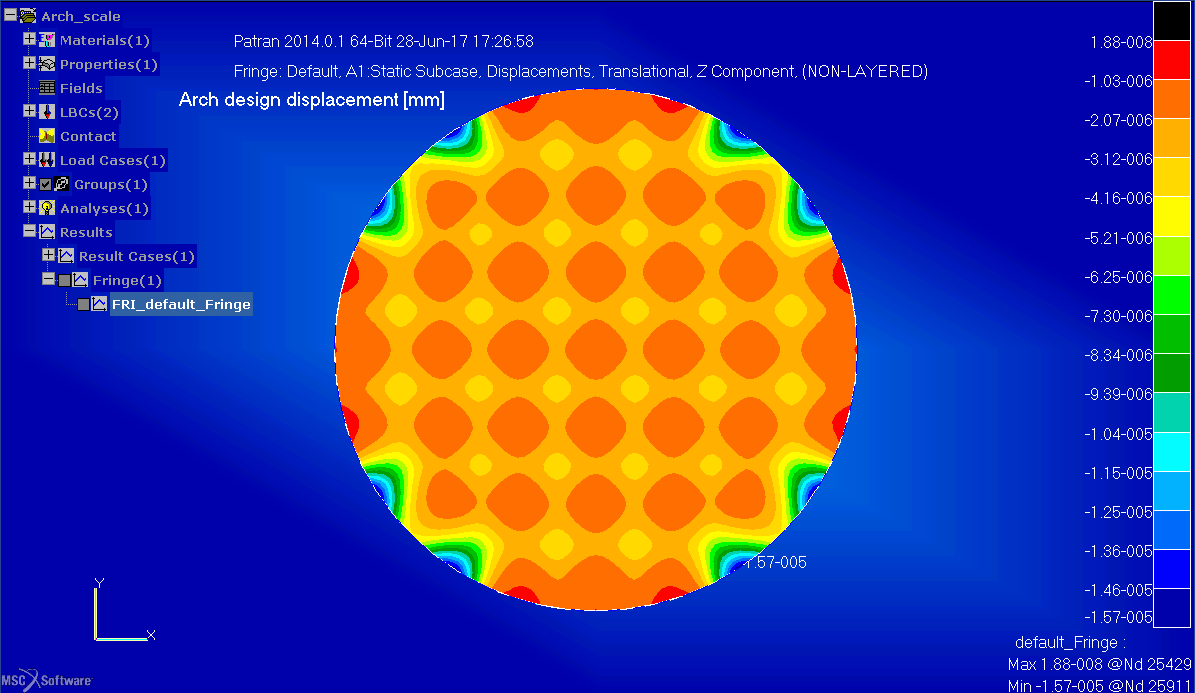}\hfill
  \includegraphics[height = 4.7cm]{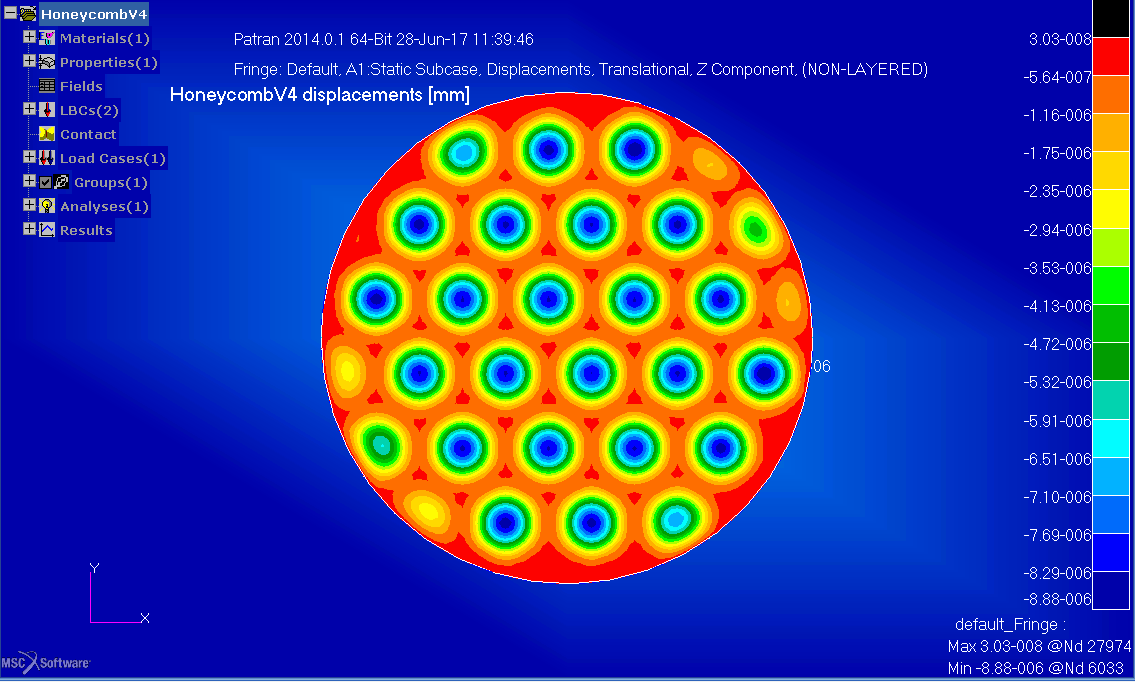}
  \caption{Comparison of the result from the simulations of Arch model (left) and honeycomb 4 model (right) in [mm]}
  \label{comparison}
\end{figure}

Figure \ref{comparison} shows the nodes displacement of the reflecting surface in the z-axis. The maximum displacements are found on the edge where there were no stiff borders and a lack of pattern. In honeycomb design this maximum displacement was -23.8nm while for the arches design it is -15.7nm. Moreover with arch design there were less displacements between the arches whereas in the honeycomb design we can see deeper node displacements. 

The arch design showed better surface quality after a polishing phase than the honeycomb. This configuration is stiffer.

\subsection{Manufacturing}

\subsubsection{3D printing method}

All the samples were manufactured with 3D printing at CA Models. There are different manufacturing methods. Below is presented the two methods used for the prototyping. 

\begin{figure}[H]
  \centering
  \includegraphics[height = 6cm]{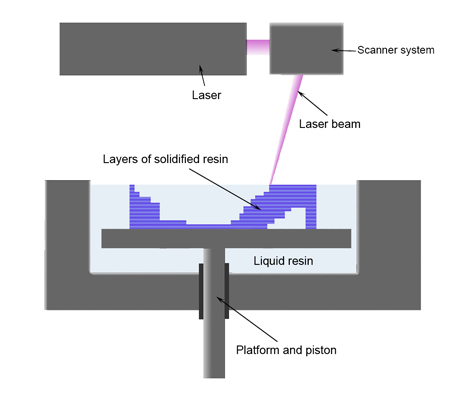}\hfill
  \includegraphics[height = 5.5cm]{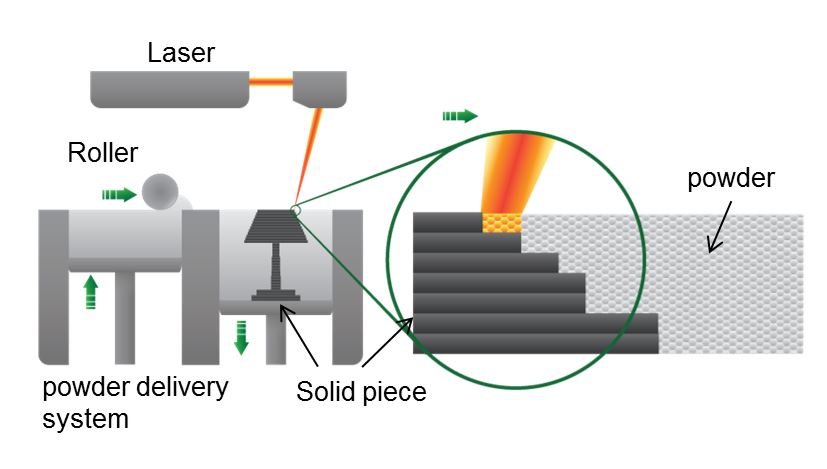}
  \caption{3D printing method, Stereo-lithography (left) and Selective laser Sintering (right) (credit: proto3000, RP Canada)}
  \label{method}
\end{figure}

Stereo-lithography (SLA) process is an additive manufacturing process, ultraviolet laser is focus on to a vat of photo-polymer resin (liquid) which is then solidified and forms a single layer of the desired 3D shape, in Figure \ref{method} (left). This process is repeated until the shape is completed. It can create 3D object in plastic and ceramic-like materials. \cite{Stampfl2014}

Selective laser Sintering (SLS) and Metal laser Sintering (MLS) are different methods to SLA. This technology uses a laser to fuse and bond small grains of material, in Figure \ref{method} (right). The laser traces the patterns of each cross section of the design onto a bed of powder. After one layer is built, the bed lowers and another layer is built on the top of the existing layers. This process continues until the 3D design is completed. It can print object in metals, polymer and composite material. \cite{Stampfl2014}

\subsubsection{Material}

\begin{figure}[H]
  \centering
  \includegraphics[scale = 0.1]{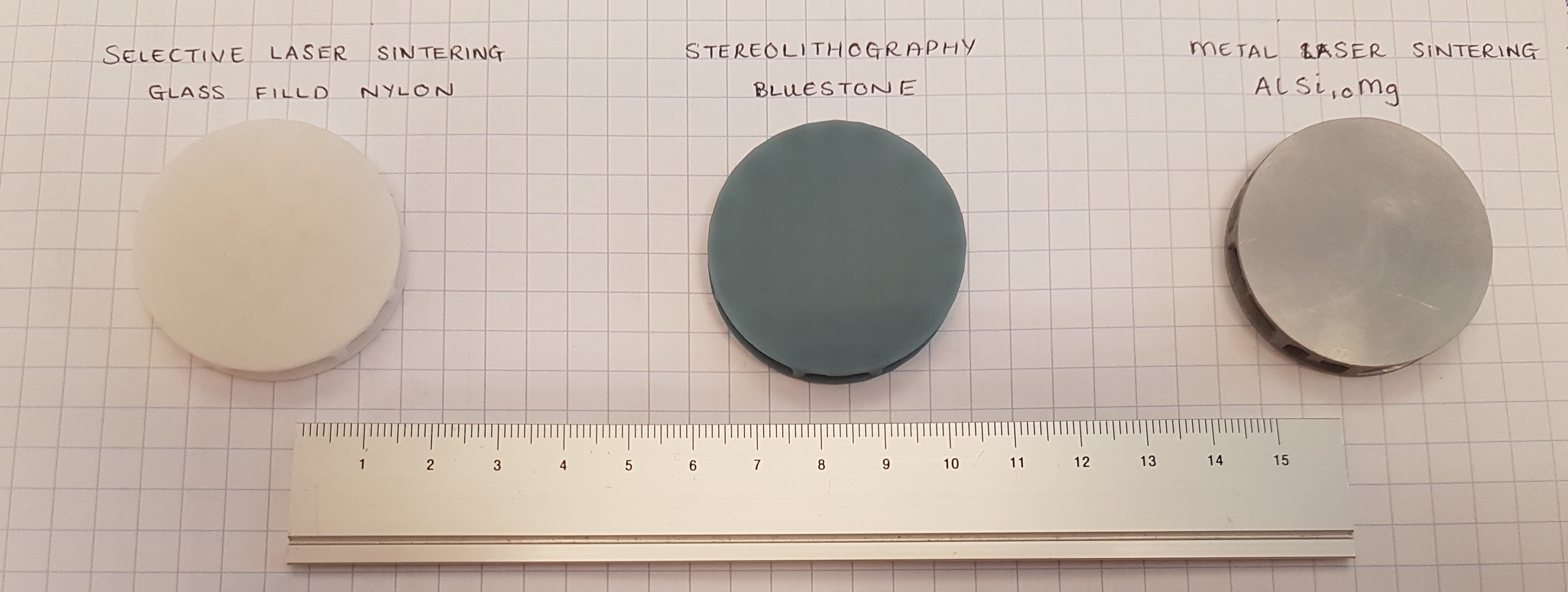}
  \caption{Prototypes picture, form left to right Glass filled nylon, Bluestone and AlSi10Mg}
  \label{samples}
\end{figure}
Figure \ref{samples} introduces the different samples. Three materials have been used, from left to right: glass filled nylon, Bluestone and aluminium (AlSi10Mg).

Glass filled nylon is a composite material. It is made by mixing polyamide powder and glass beads. The surface of the material is white and slightly porous. Compared to native polymer, glass filled improves mechanical properties of rigidity and strength by reduced the non-lubricated wear properties. This material is largely used in 3D printing with Selective Laser Sintering. \cite{chung2006processing}

Bluestone is a trade name for a ceramic-like material manufactured with stereo-lithography process. This material is not used in optics. It is a new composite material resulting of the development of 3D printing. Its Young's modulus is 7.6 GPa and this material shows exceptional stiffness. \cite{CAModelsBlu}

AlSi10Mg is aluminium alloy. AlSi10Mg is printing with MLS which is similar process than SLS. It offers good strength, hardness and dynamic properties. Its Young's modulus is around 70 GPa. This material has excellent machinability so we could apply polishing on its surface to reduce the roughness. \cite{CAModelsAl}

\subsubsection{Post processing}

The initial goal to evaluate the surface quality of the samples immediately post-print and after some basic smoothing processes in order to quantify the surface quality prior to grinding and polishing. Samples were processed at different levels, the sanded sample was sanded by hand, the skim sample was sanded with finer grit and the blasted samples was bead blasted. Glass filled nylon is not strong enough to be post polished so there is only the raw sample. Table \ref{resumesamples} summarises the samples and post-processing treatments. Checked boxes mean that the sample has been done. 

\begin{table}[H]
\begin{center}
\begin{tabular}{|c|c|c|c|c|c|c|}
  \hline
  Material/Post polishing & Raw & Sanded & Skim & Blasted & Pure Ni coating & NiP coating \\
  \hline
  AlSi10Mg & X &   &  & X  &    & X\\
  \hline
  Bluestone & X & X & X & X & X &   \\
  \hline
  Glass filled Nylon & X&  &  &  & X &  \\
  \hline
\end{tabular}
\end{center}
\caption{Summary of the samples}
  \label{resumesamples}
\end{table}

From the raw samples of each material we added a step of coating. The coating is added to provide a mechanically harder surface to polish to ensure good reflective properties. Bluestone and glass filled nylon were coated with pure nickel and the AlSi10Mg with nickel phosphor (NiP), as shown in Table\ref{resumesamples}. Nickel phosphor is commonly used in X-ray optics manufacture due to its very good polishing properties. The coating process for nickel phosphor is made by a catalytic electroless process at high temperature and for that reason it was not applied to the glass filled nylon and Bluestone samples. Indeed their mechanical properties change when material reach high temperature. Only Bluestone and glass filled nylon the samples are coating with pure nickel, at room temperature, which has also good properties in polishing. Pure nickel and nickel phosphor have roughly the same density so both coatings have similar weight.

\subsection{Surface quality}

Only non destructive measurements have been performed on the samples \cite{SPIEpaper}. We measured the surface profile, the roughness and waviness.

\subsubsection{Surface profilometry}

\begin{figure}[H]
  \centering
  \includegraphics[width = 0.45\linewidth]{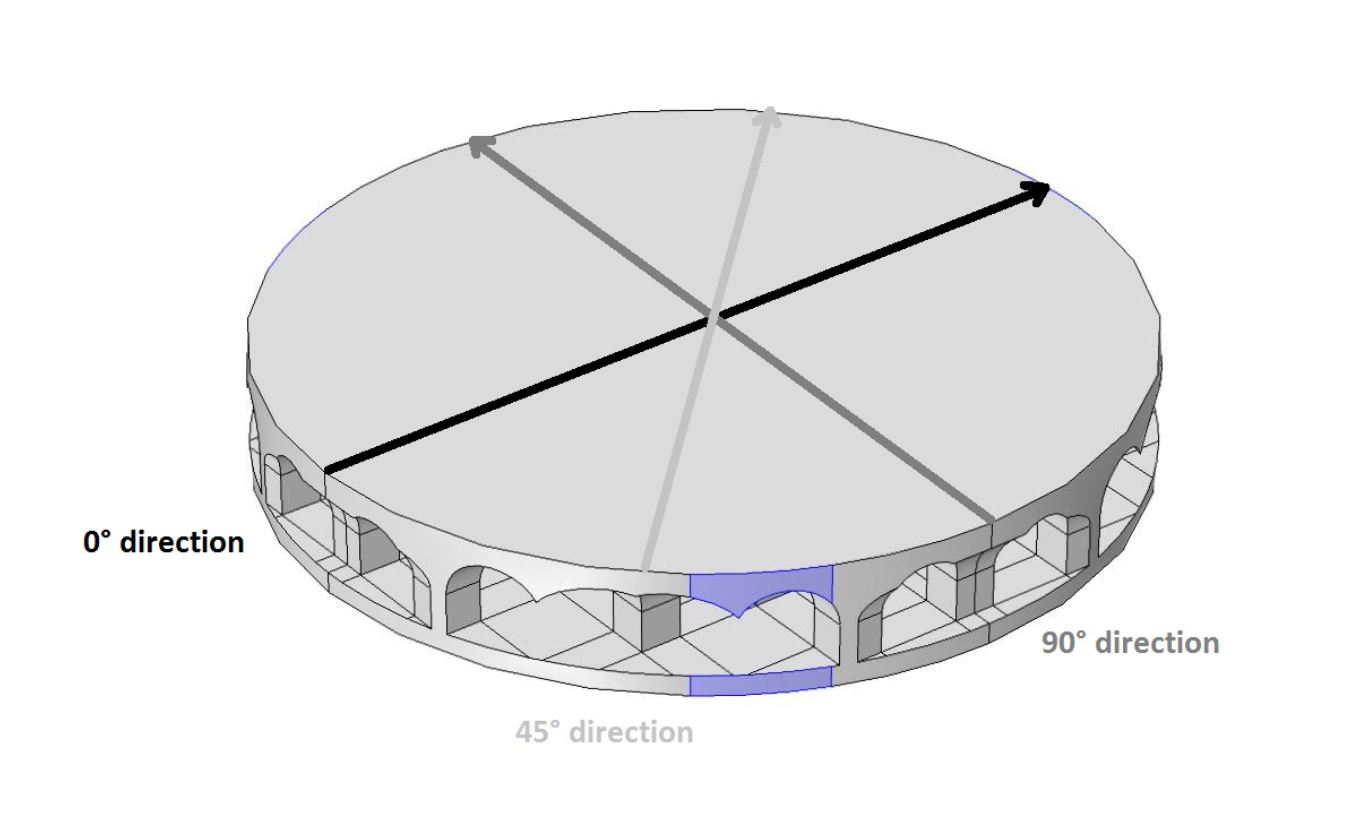}\hfill
  \includegraphics[height = 5cm]{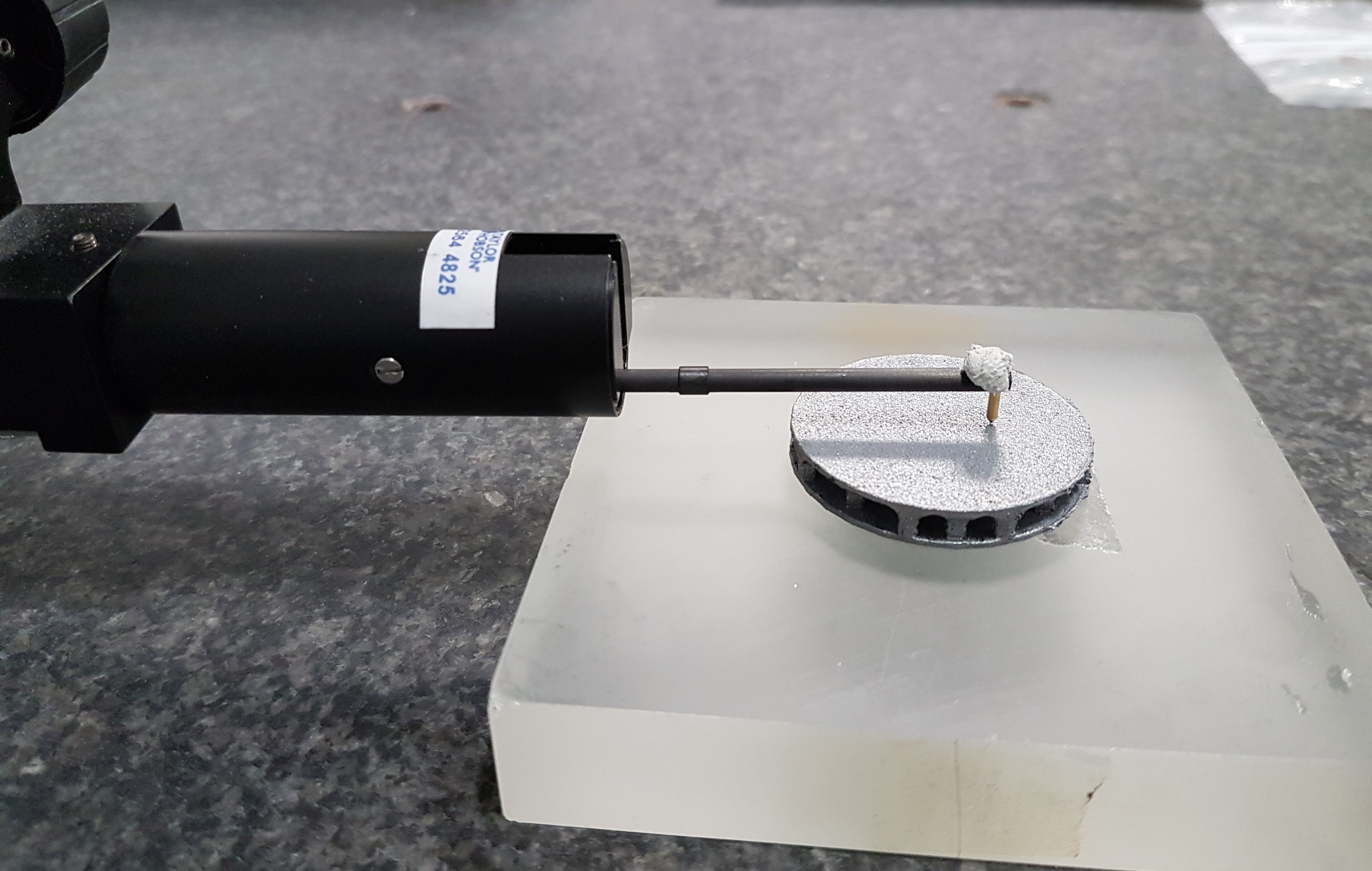}
  \caption{CAD view of the prototype design with the three direction measurements (left), Picture of the profilometry machine working on the AlSi10Mg raw sample (right)}
  \label{surfacemeasurement}
\end{figure}

A Taylor Hobson Form Talysurf Intra measured the height of the surface at given locations along a line, in Figure \ref{surfacemeasurement} (right). To sample the whole surface measurements were taken in 3 different directions, in Figure \ref{surfacemeasurement} (left), 0$^{\circ}$, 45$^{\circ}$ and 90$^{\circ}$ directions. The measurement was performed three times for each direction in order to have more accuracy in the surface profile.

\begin{figure}[H]
  \centering
  \includegraphics[width = 1\linewidth]{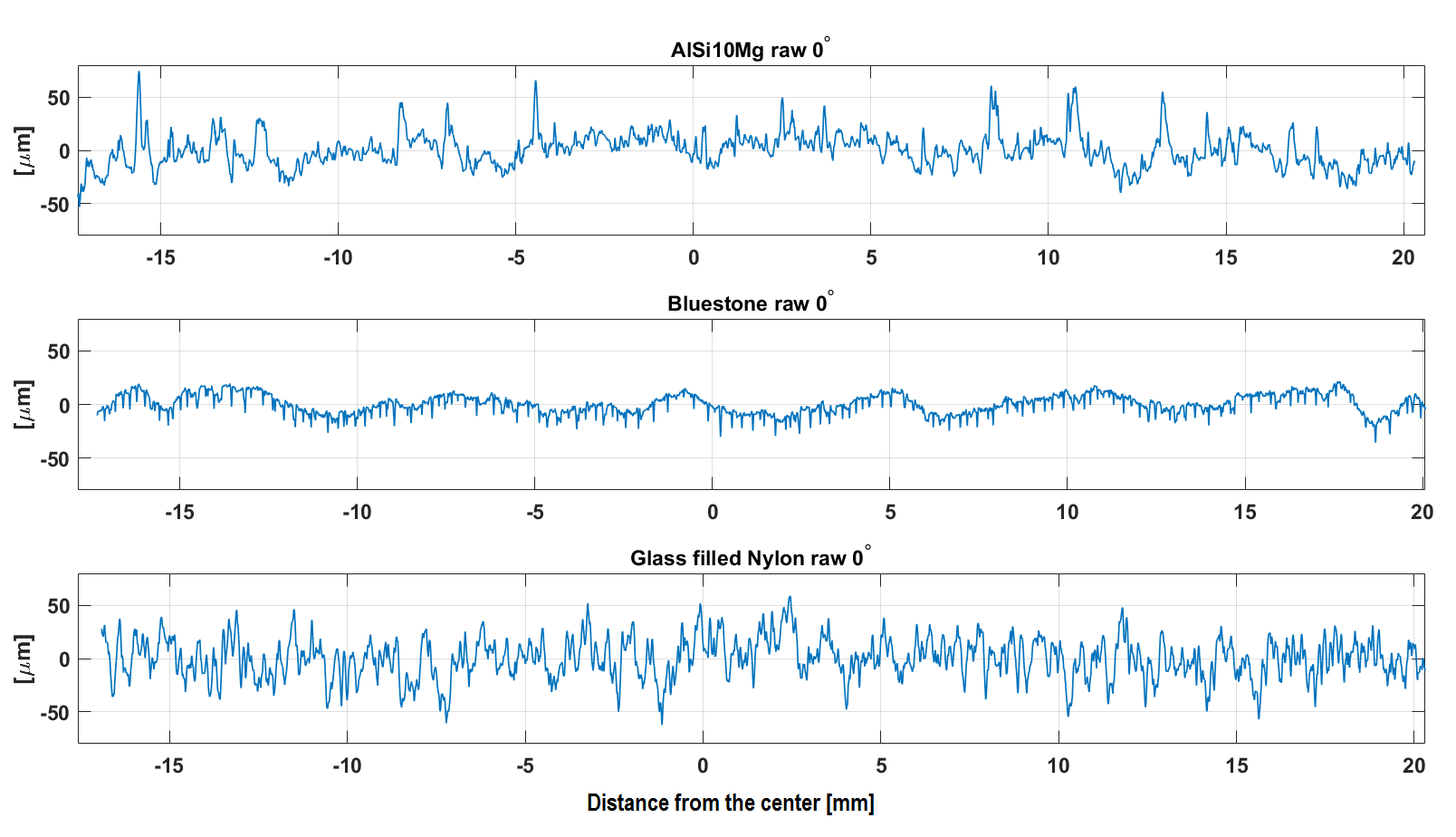}
  \caption{Average surface profile for raw samples on 0$^{\circ}$ direction for AlSi10Mg (top), Bluestone (middle), Glass filled Nylon (bottom)}
  \label{profil}
\end{figure}

Figure \ref{profil} shows the average surface profile for AlSi10Mg, Bluestone and glass filled nylon raw sample in the 0$^{\circ}$ direction. The amplitude of the surface profile varies depending of the materials. Glass filled nylon surface is a lot rougher than AlSi10Mg and Bluestone. Furthermore on these last one we observed waviness pattern, which depicts approximately 6 peaks. This is a consequence of their orientation within the 3D printer and the arches lightweight structure.

\subsubsection{Fourier analysis}

In order to remove the waviness pattern from previous data, Fourier analysis was applied. The Fourier decomposition is a way to represent a function as the sum of simple sine waves. It is used in signal treatment to isolate narrow band component of a compound waveform. 

\begin{figure}[H]
  \centering
  \begin{minipage}[l]{0.79\linewidth}
  \includegraphics[width = 1\linewidth]{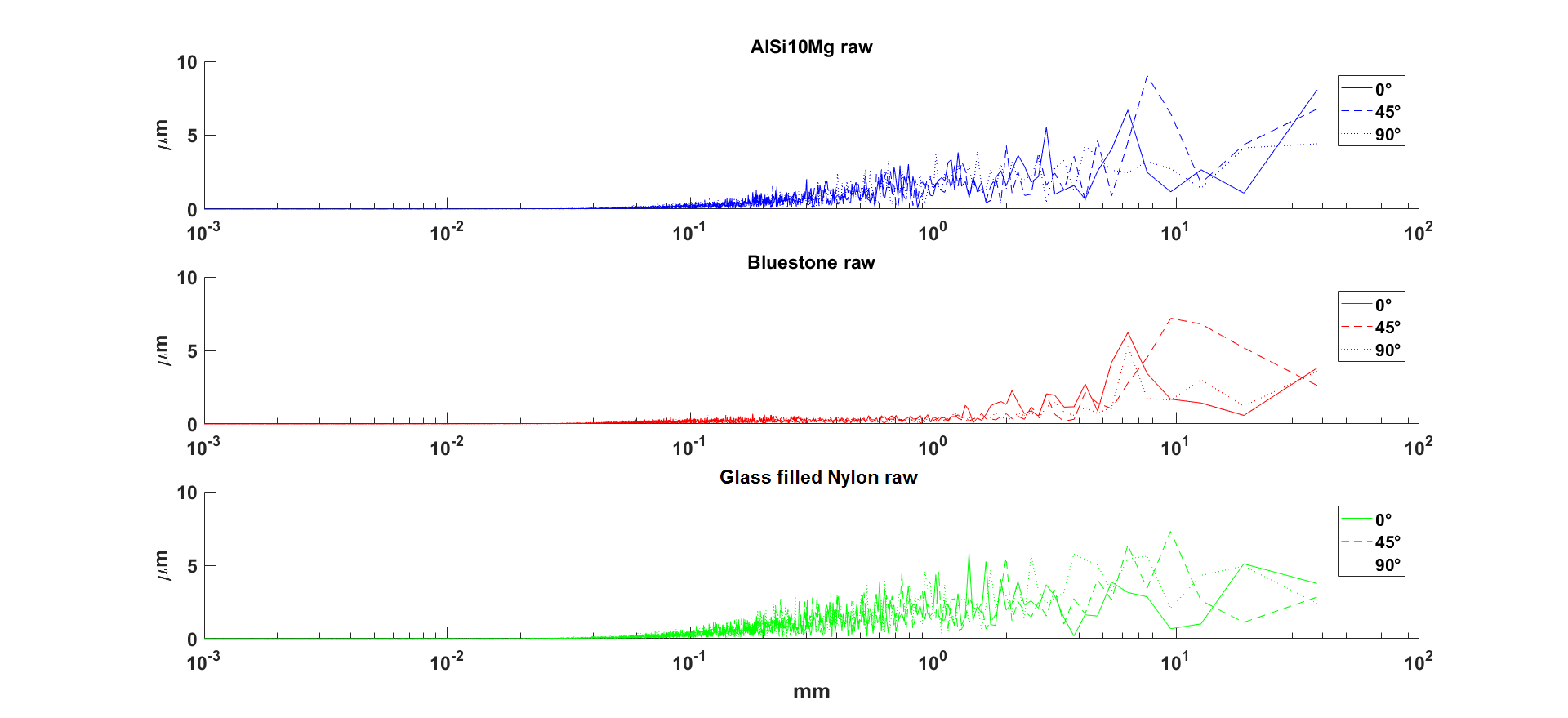}
  \end{minipage}
  \begin{minipage}[r]{0.2\linewidth}
  \includegraphics[width = 1\linewidth]{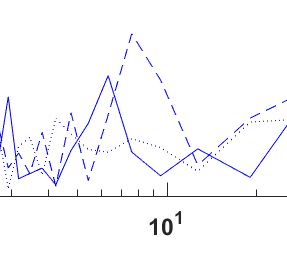}\\
  \includegraphics[width = 1\linewidth]{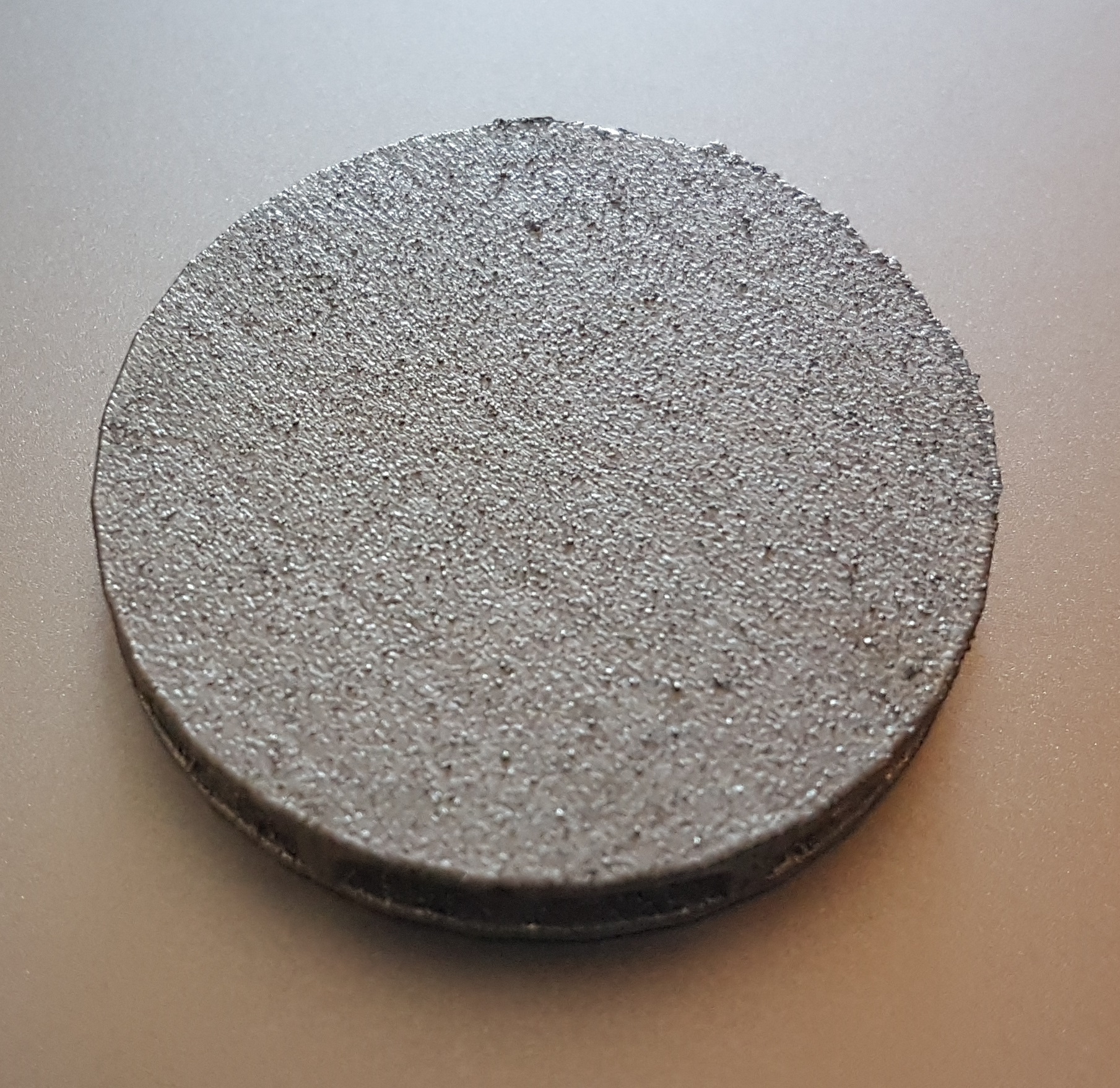}
  \end{minipage}
  \caption{Fourier analysis from the profilometry data. Spatial frequency for raw samples (left) in AlSi10Mg (top), Bluestone (middle) and Glassfilled nylon (bottom), Zoom on low frequencies for AlSi10Mg raw (top right) Picture of AlSi10Mg raw sample surface (low right)}
  \label{fourier}
\end{figure}

Figure \ref{fourier} (left) depicts the spatial frequencies for AlSi10Mg raw, Bluestone raw and glass filled nylon raw. The top right picture in Figure \ref{fourier} is a zoom of the AlSi10Mg raw sample for low frequencies. There is two obvious peaks which depicts the waviness pattern explained in section 3.3.1. The larger bump were in the diagonal (45$^{\circ}$) direction and in the vertical direction (0$^{\circ}$) the bumps had slightly lower magnitude. On Bluestone spatial frequencies these patterns can be seen in all directions. The bumps magnitude and period were similar, which obviously leads to a diagonal pattern with larger bumps. As we expected from the surface profile measurement there were no evident pattern for the Glass filled nylon.  

In order to only investigate the roughness we used the ISO standard 4288-1996 definition. \cite{2003exploring}
This standard gives for an anticipated roughness the necessary cut-off to highlight the roughness, in our case the cut off is between 0.8mm and 2.5mm. Considering these filters, roughness is below 1mm-2mm, waviness between 2mm-10mm and form is above 10mm. These are approximate values and they vary between measurement length and expected roughness.

\subsubsection{Discussion}
For all the samples and directions we calculate the Root Mean Square (RMS) and Peak to Valley (PV) values \cite{2003exploring} to compare the surface quality of the samples.

\begin{figure}[H]
  \centering
  \includegraphics[width = 1\linewidth]{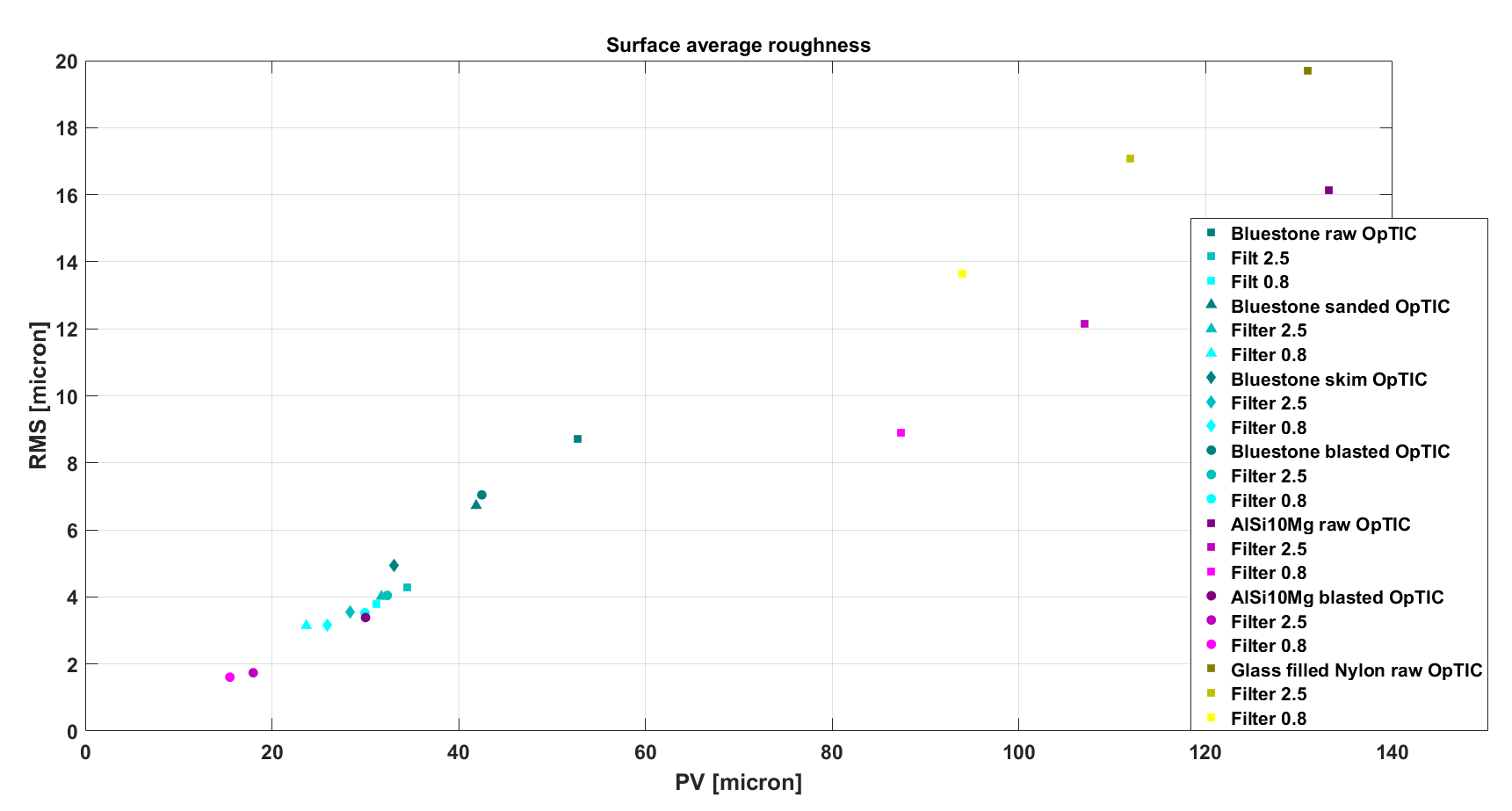}
  \caption{Summary of PV and RMS in microns for all the samples without filter, with 2.5 filter and with 0.8 filter}
  \label{resume}
\end{figure} 
In Figure \ref{resume} is presented the RMS in function of PV for each sample, without filter, with 2.5mm filter and 0.8mm filter. We take the mean of RMS and PV of the three orientation to obtain only one value per sample. In blue there are all the Bluestone samples, in purple all the AlSi10Mg samples and in yellow the glass filled nylon sample. Then in the graph the post polishing process are also highlighted, the raw sample are squared, the sanded is triangular, the skim is diamond and  the blasted are circular. Finally to show the impact of the filter there is a colour gradient, the lighter is the smaller filter and the darker is without filter. 

As expected the raw samples without filter had the highest roughness. Particularly for glass filled nylon and AlSi10Mg, besides for these samples the filtering cut-off shows a lot of surface shape and waviness. From Figure \ref{resume} the gap between the non filtered and filtered data was huge, even between 2.5mm cut off and 0.8mm which attests the existence of a deep waviness pattern. While in Bluestone the first filter deleted the majority of the waviness pattern. Bluestone showed the best surface quality after printing without any post-processing.

The AlSi10Mg post-processed sample showed very good surface quality, RMS under 4$\mu$m, the filtering showed there is still some waviness pattern after the processing but not to much because 2,5mm and 0.8mm cut were close. Blasted sample is 5 times better than raw in terms of PV and RMS. Finally, in comparison to all the samples AlSi10Mg blasted had the best surface in terms of roughness for very little effort in terms of post-processing.

For the post-process Bluestone samples  the surface quality is better than the raw samples as expected. After applying the filter the waviness pattern is completely remove and all the data are located in a square between 20$\mu$m and 40$\mu$m of PV and 3$\mu$m-5$\mu$m of RMS so Bluestone shows very good result in term of roughness.

\subsection{Topology optimisation}

Topology optimisation is a mathematical method which optimises the design for a given set of loads, boundary conditions and constraints with the goal of maximising the performance of the system on pre-defined parameter, in our case the volume and so the weight. \cite{bendsoe2004}

\begin{figure}[H]
  \centering
  \includegraphics[width = 0.5\linewidth]{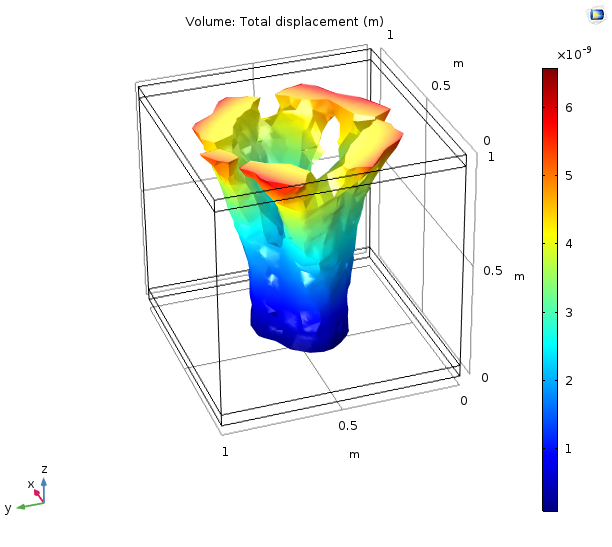}\hfill
  \includegraphics[width = 0.5\linewidth]{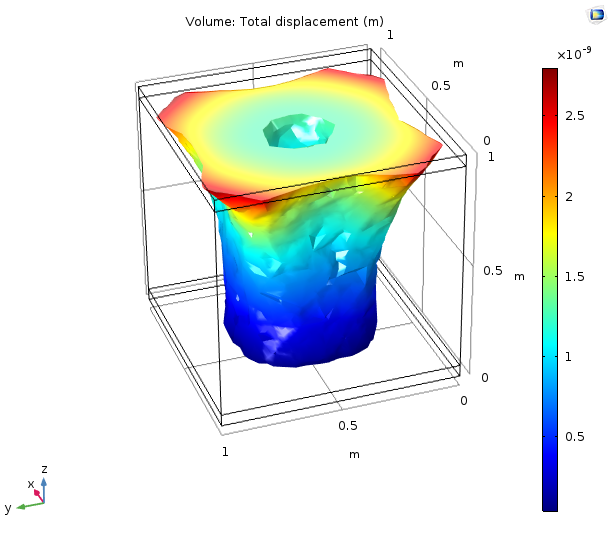} 
  \caption{Optimum design given by COMSOL software after topology optimisation of the volume on a cube for 30\% keeping mass (left) and 50\% keeping mass (right)}
  \label{topo2}
\end{figure} 

Figure \ref{topo2} shows the result of topology optimisation for a cube sandwiched between two plates. Aluminium was used in the simulation, the base was fixed and a uniform force was applied on the top. In Figure \ref{topo2} (left) there is the result for 30\%  mass kept, that involves to have a large hole in the centre of the column and opening on the edges. For the model with 50\%  mass kept we obtained a small hole in the centre of the column but this design could not be printed in 3D. Indeed 3D printing can't print internal holes, the process needs to add support and so the support needs to be removed afterwards. With the model 30\% as there were opening on the edges of the column the interior support could be removed. 

This innovative technology will be used to design the next set of samples. We should take in account that all the solution from topology optimisation cannot be printed, so we should adapt the design to the requirements of 3D printing.

\section{Conclusion and perspectives}

In the first section of this paper, we demonstrated that we can manufacture off axis parabolas using stress polishing and 3D printing. We developed a new fabrication process simpler and faster than the previous one for space mirrors keeping the quality on the surface.

In the second part of this paper, surface quality of lightweight optics manufacture by 3D printing was investigated. Bluestone is desirable due to its smooth surface post-print and low density; however the high CTE would make it undesirable for environments with large temperature gradients. AlSi10Mg showed very good result after post processing.

The surface quality of of the polished coated samples was presented by Atkins et al in SPIE San Diego 2017 \cite{SPIEpaper}. All the polished samples are good enough for visible/optical wavelengths which is much better than required for infrared. Aluminium NiP would be classified as near high surface quality. They are not good enough for X-ray optic, so we will continue to develop and investigate the polishing treatment and the design. In the future we use topology optimisation computation to improve the design. 

\section{ACKNOWLEDGEMENTS}
The authors would like to acknowledge the European commission for funding this work through the Program H2020-ERC-STG-2015 – 678777 of the European Research Council. In addition, part of this project has been funded by the UK Space Agency's National Space Technology Programme and C. Atkins acknowledges the following with respect to her research fellowship: this project has received funding from the European Union’s Horizon 2020 research and innovation programme under the Marie Skłodowska-Curie grant agreement (GA) No. 665593 awarded to the Science and Technology Facilities Council.

\bibliography{report} 

\begin{thebibliography}{10}

\bibitem{Jane2012}
Bird, J., ``Exploring the 3d printing opportunity,'' {\em Financial Times}
  (August 2012).

\bibitem{Enrico2018}
Hilpert, E., Hartung, J., Risse, S., Eberhardt, R., and Tddotunnermann, A.,
  ``Precision manufacturing of a lightweight mirror body made by selective
  laser melting,'' (2018).

\bibitem{SPIEpaper}
Atkins, C., Feldman, C., Brooks, D., Watson, S., Cochrane, W., Roulet, M.,
  Doel, P., Willingale, R., and Hugot, E., ``Additive manufactured x-ray optics
  for astronomy,'' {\em Proc.SPIE}~{\bf 10399},  10399 -- 10399 -- 15 (2017).

\bibitem{2015arXiv150303757S}
{Spergel}, D., {Gehrels}, N., and {Baltay}, e.~a., ``{Wide-Field InfrarRed
  Survey Telescope-Astrophysics Focused Telescope Assets WFIRST-AFTA 2015
  Report},'' {\em ArXiv e-prints}  (Mar. 2015).

\bibitem{reflynx}
Gaskin, J.~A., Allured, R., Bandler, S.~R., Basso, S., and al., ``Lynx mission
  concept status,'' {\em Proc. SPIE 10397, UV, X-Ray, and Gamma-Ray Space
  Instrumentation for Astronomy XX, 103970S} ,  13 (Sept 2017).

\bibitem{Stephen2016}
O'Dell, S.~L., Allured, R., Ames, A.~O., and al., ``Toward large-area
  sub-arcsecond x-ray telescopes ii,'' {\em Proc. SPIE 9965, Adaptive X-Ray
  Optics IV, 996507} ,  17 (Oct 2016).

\bibitem{Noll:76}
Noll, R.~J., ``Zernike polynomials and atmospheric turbulence$\ast$,'' {\em J.
  Opt. Soc. Am.}~{\bf 66},  207--211 (Mar 1976).

\bibitem{Lemaitre:72}
Lema\^{i}tre, G., ``New procedure for making schmidt corrector plates,'' {\em
  Appl. Opt.}~{\bf 11},  1630--1636 (Jul 1972).

\bibitem{lemaitre2008astronomical}
Lemaitre, G.,  [{\em Astronomical Optics and Elasticity Theory: Active Optics
  Methods}{\nolinebreak\hspace{0.1em}]}, Astronomy and Astrophysics Library,
  Springer Berlin Heidelberg (2008).

\bibitem{Hugot2008}
Hugot, E., Laslandes, M., Ferrari, M., Vives, S., Moindrot, S., Hadi, K.~E.,
  and Dohlen, K., ``Active optics: off axis aspherics generation for high
  contrast imaging,'' {\em Proc.SPIE}~{\bf 10565},  10565 -- 10565 -- 7 (2018).

\bibitem{Schott263}
SCHOTT, A., ``Thermal expansion of zerodur,'' tech. rep. (May 2013).

\bibitem{hugot:tel-00519452}
Hugot, E., {\em {Astronomical optics and elasticity - Active Optics for future
  Extremely Large Telescopes and their instrumentation}}, theses,
  {Universit{\'e} de Provence - Aix-Marseille I} (Oct. 2007).

\bibitem{Chandra2012}
Weisskopf, M.~C., ``Chandra x-ray optics,'' {\em Optical Engineering}~{\bf 51},
   51 -- 51 -- 8 (2012).

\bibitem{Stampfl2014}
Stampfl, J. and Hatzenbichler, M.,  [{\em Additive Manufacturing
  Technologies}{\nolinebreak\hspace{0.1em}]},  20--27, Springer Berlin
  Heidelberg, Berlin, Heidelberg (2014).

\bibitem{chung2006processing}
Chung, H. and Das, S., ``Processing and properties of glass bead
  particulate-filled functionally graded nylon-11 composites produced by
  selective laser sintering,'' {\em Materials Science and Engineering: A}~{\bf
  437}(2),  226--234 (2006).

\bibitem{CAModelsBlu}
CAModels, ``{Data sheet - Bluestone}.''
  http://www.camodels.co.uk/media/1253/00-sla-bluestone-ready.pdf.

\bibitem{CAModelsAl}
CAModels, ``{Data sheet - AlSi10Mg}.''
  http://www.camodels.co.uk/media/1288/metal-ad-alsi10mg.pdf.

\bibitem{2003exploring}
Hobson, T., ``Exploring surface texture: A fundamental guide to the measurement
  of surface finish,'' tech. rep. (2003).

\bibitem{bendsoe2004}
Bendsøe, M.~P. and Sigmund, O.,  [{\em Topology Optimization Theory, Methods,
  and Applications}{\nolinebreak\hspace{0.1em}]}, Springer-Verlag Berlin
  Heidelberg, Berlin, Heidelberg (2004).

\end{thebibliography}
\bibliographystyle{spiebib} 

\end{document}